\documentclass[revtex4,useAMS,usenatbib]{emulateapj}

\usepackage{amsmath}
\usepackage{amsfonts}
\usepackage{amssymb}
\usepackage[usenames,dvipsnames]{xcolor}
\usepackage{graphicx}
\usepackage{epstopdf}
\usepackage{epsfig}
\DeclareGraphicsExtensions{.pdf,.png,.jpg}
\graphicspath{{.}}
\usepackage{tabularx}
\usepackage{rotating}
\usepackage{lscape}
\usepackage{bold-extra}
\usepackage{appendix}
\usepackage{listings}
\usepackage{float}
\usepackage{pdfpages}
\usepackage{natbib}
\def\apjl{ApJL}

\def\apjs{ApJS}

\def\aap{A\&A}

\def\micron{$\mu$m }

\def\Msun{M$_{\odot}$}
\def\7C{7C\,1753$+$6311}
\def\ClG7C{CARLA\,J1753$+$6311}
\usepackage{url}
\usepackage{hyperref}
\usepackage{enumerate}

\let\oldsqrt\sqrt
\def\sqrt{\mathpalette\DHLhksqrt}
\def\DHLhksqrt#1#2{%
\setbox0=\hbox{$#1\oldsqrt{#2\,}$}\dimen0=\ht0
\advance\dimen0-0.2\ht0
\setbox2=\hbox{\vrule height\ht0 depth -\dimen0}%
{\box0\lower0.4pt\box2}}

\slugcomment{}

\shorttitle{}
\shortauthors{Cooke et al.}

\begin{document}


\title{A mature galaxy cluster at $z=1.58$ around the radio galaxy 7C\,1753$+$6311}


\author{E. A. Cooke$^{1}$, N. A. Hatch$^{1}$, D. Stern$^{2}$, A. Rettura$^{3}$, M. Brodwin$^{4}$, A. Galametz$^{5}$, D. Wylezalek$^{6}$, C. Bridge$^{2,7}$, C. J. Conselice$^{1}$, C. De Breuck$^{8}$, A. H. Gonzalez$^{9}$, M. Jarvis$^{10,11}$}
\affil{$^{1}$School of Physics and Astronomy, University of Nottingham, University Park, Nottingham NG7 2RD, UK\\
$^{2}$Jet Propulsion Laboratory, California Institute of Technology, Pasadena, CA 91109, USA\\
$^{3}$Infrared Processing and Analysis Center, California Institute of Technology, Pasadena, CA 91125, USA\\
$^{4}$Department of Physics and Astronomy, University of Missouri, Kansas City, MO 64110, USA\\
$^{5}$Max-Planck-Institut fuer Extraterrestrische Physik, Giessenbachstrasse, D-85748 Garching, Germany\\
$^{6}$Department of Physics and Astronomy, Johns Hopkins University, 3400 N. Charles St, Baltimore, MD 21218, USA\\
$^{7}$California Institute of Technology, 1200 E. California Blvd., Pasadena, CA 91125, USA\\
$^{8}$European Southern Observatory, Karl Schwarzschild Strasse 2, 85748 Garching bei Munchen, Germany\\
$^{9}$Department of Astronomy, University of Florida, Gainesville, FL 32611, USA\\
$^{10}$Astrophysics, University of Oxford, Denys Wilkinson Building, Keble Road, Oxford OX1 3RH, UK\\
$^{11}$Physics Department, University of the Western Cape, Bellville, South Africa\\
}
\email{Elizabeth.Cooke@nottingham.ac.uk}

%
%




\begin{abstract}
We report on the discovery of a $z=1.58$ mature cluster around the high-redshift radio galaxy 7C\,1753$+$6311, first identified in the Clusters Around Radio-Loud AGN survey. Two-thirds of the excess galaxies within the central 1\,Mpc lie on a red sequence with a colour that is consistent with an average formation redshift of $z_f\sim3$. We show that $80\pm6$\% of the red sequence galaxies in the cluster core are quiescent, while the remaining $20\%$ are red due to dusty star formation. We demonstrate that the cluster has an enhanced quiescent galaxy fraction that is three times that of the control field. We also show that this enhancement is mass dependent: $91\pm9$\% of the $M_* >10^{10.5}$\,M$_{\odot}$ cluster galaxies are quiescent, compared to only $36\pm2$\% of field galaxies, whereas the fraction of quiescent galaxies with lower masses is the same in the cluster and field environments. 
The presence of a dense core and a well-formed, quiescent red sequence suggest that this is a mature cluster. This means that distant radio galaxies do not solely reside in young, uncollapsed protoclusters, rather they can be found in clusters in a wide range of evolutionary states. 

\end{abstract}




\section{Introduction}
To understand the formation history of galaxies in the densest environments, clusters, it is important to study their progenitors: high-redshift protocluster galaxies. 
Several techniques are now employed to locate clusters and protoclusters at $z>1$, such as large photometric surveys \citep[e.g.][]{Chiang2014,Stanford2014}, surveys exploiting the Sunyaev Zel'dovich (SZ) effect \citep[e.g.][]{Hasselfield2013,Bleem2015,Planck2015}, and X-ray detections of the intracluster medium (ICM) \citep[e.g.][]{Willis2013}. 
Unfortunately, many of these methods are expensive, requiring deep coverage of large fields-of-view in order to locate the rare overdensities. 
Massive clusters have been found with the SZ effect out to $z\sim1.5$, with a few candidates at even higher redshifts \citep[e.g.][see also \citealp{Brodwin2012}]{Tozzi2015}. These, however, are rare systems; SZ and X-ray surveys struggle to find the more typical, lower mass clusters at $z>1.5$ since the signal from these methods scales with cluster mass. 
The high-redshift progenitors of the majority of local $M \sim 10^{14}$\,\Msun\ clusters will be missed since they lack sufficiently massive cluster cores at $z\gtrsim1.5$ to be detectable with current instruments \citep{Chiang2013,Muldrew2015}. 
A targeted approach reduces the strain on telescope time and can pinpoint clusters even at the highest redshifts. However, the right beacon needs to be used.

Radio-loud active galactic nuclei (RLAGN) preferentially reside in dense environments at high redshift \citep[e.g.][]{Galametz2012, Wylezalek2013}, which are significantly denser than the environments of radio-quiet galaxies of the same stellar mass \citep{Hatch2014}, which is also predicted from models \citep{Orsi2015}. These RLAGN provide one of the most efficient ways to find and study large-scale structure formation, particularly at the highest redshifts. However, if RLAGN preferentially reside in clusters of a certain age or stage of collapse, then our view of cluster formation will be biased. For example, \citet{Simpson2002} and \citet{vanBreukelen2009} suggest distant radio galaxies pinpoint merging clusters. 
 
Most confirmed cluster progenitors, known as protoclusters, have been identified with Ly$\alpha$ emitters, H$\alpha$ emitters, or Lyman-break galaxies, which are tracers of young systems \citep[e.g.][]{Overzier2005,Venemans2007,Cooke2014}. So the methods by which protoclusters have been identified preferentially pinpoint young, forming galaxies, and clusters that contain mature, passively evolving galaxies may be missed.

By contrast, large cluster surveys using \emph{Spitzer} data are not intrinsically biased to star-forming galaxies \citep{Simpson1999,Eisenhardt2008,Muzzin2009,Galametz2012,Wylezalek2013,Rettura2014}. The criterion proposed by \citet{Papovich2008} uses a 3.6\,\micron $-$ 4.5\,\micron colour selection to isolate all types of galaxies at $z>1.3$, thanks to the peak of stellar emission near 1.6\,\micron in galaxy spectral energy distributions (SEDs) moving into these bands at $z>1$. 
This feature is present in galaxies of all types so selecting only on this criteria does not greatly bias the galaxy selection.

One very successful \emph{Spitzer} survey is the Clusters Around Radio-Loud AGN survey \citep[CARLA;][]{Wylezalek2013,Wylezalek2014}, which imaged over 400 fields surrounding RLAGN, and identified $\sim200$ cluster and protocluster candidates at $1.3<z<3.2$. Using the \emph{Spitzer} wavebands to select both star forming and passively evolving cluster members, these clusters can be used to investigate whether RLAGN are biased tracers of clusters that reside preferentially in younger structures, or whether the young structures discovered to date are due to the protocluster confirmation techniques used. 

\begin{figure}
\centering
\includegraphics[trim = 0mm -3mm 0mm 0mm, clip,width=\columnwidth]{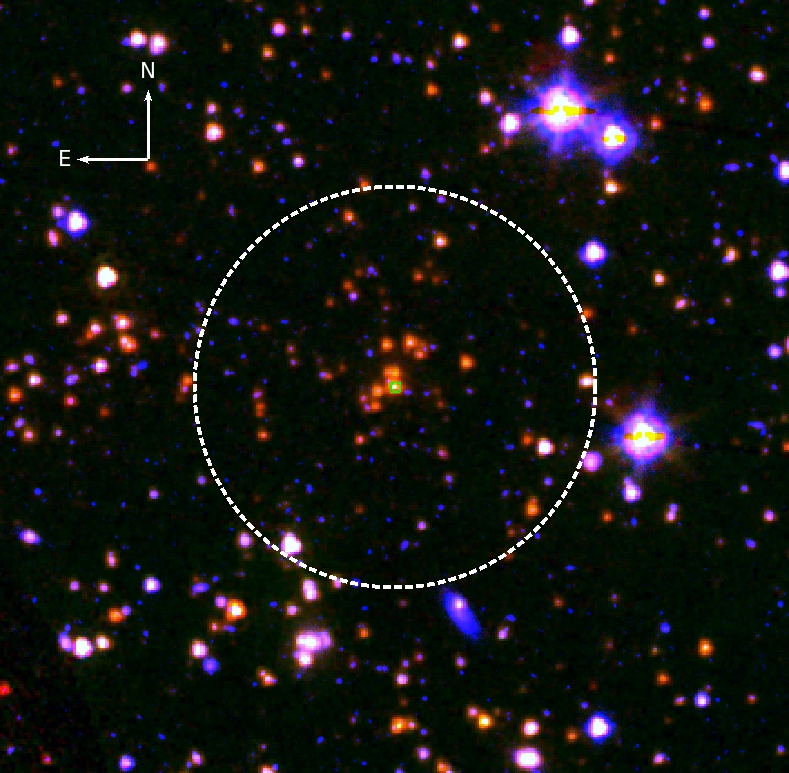}
\caption{$i',[3.6],[4.5]$ three-colour image of the field around \7C. The central RLAGN is marked with a green square. There are several red sources clustered around the RLAGN. 
The white dashed circle shows a 0.9\,arcmin radius around the RLAGN. 0.9\,arcmin at $z=1.58$ corresponds to 0.46\,Mpc in physical coordinates. }
\label{fig:fieldimg}
\end{figure}

Based on deep optical imaging, \citet{Cooke2015} identified a subset of the densest CARLA fields that show signs of being mature clusters, exhibiting red sequences, and dense cores of red galaxies. 
Here we examine one of these fields, around the RLAGN \7C (Figure \ref{fig:fieldimg}). \citet{Lacy1999} tentatively assigned a redshift to this galaxy of $z=1.95$ based on an uncertain emission line at 4854\AA\ assumed to be \ion{He}{2}~$\lambda 1640$ and the possible detection of an associated Ly$\alpha$ break. This redshift was assigned a quality ``$\gamma$'', indicating an ``uncertain'' redshift. Here we report the first robust spectroscopic redshift for \7C, confirming it to instead be at $z=1.58$, and examine the surrounding cluster environment. 
Section \ref{sec:method} outlines our data and methods used. In Section \ref{sec:spectrum} we present a new, deep, optical spectrum of \7C which confirms its redshift as being $z=1.58$. Section \ref{sec:results} then investigates the properties of the galaxies surrounding \7C, and Section \ref{sec:conclusion} presents our conclusions.

In the following, all magnitudes and colours are in the AB photometric system and we assume a $\Lambda$CDM cosmology with $H_0=70$\,km\,s$^{-1}$\,Mpc$^{-1}$, $\Omega_m = 0.3$, and $\Omega_{\Lambda} =0.7$.

\section{Data and Method}
\label{sec:method} 

\subsection{Imaging}
The field surrounding \7C was imaged with \emph{Spitzer}'s Infrared Array Camera \citep[IRAC;][]{Fazio2004} at 3.6\,\micron and 4.5\,\micron by the CARLA survey \citep{Wylezalek2013}, reaching a $3\sigma$ depth of $23.8$\,mag and $24.4$\,mag  at 3.6\,\micron and 4.5\,$\mu$m, respectively. 
This field was identified as a protocluster candidate with a $4.5\sigma$ overdensity by \citet{Wylezalek2013} and was followed up in $i'$ and $J$ using the William Herschel Telescope in La Palma. The $i'$ band image was taken with the auxiliary-port camera (ACAM), with an exposure time of 6000\,s. Full details of the $i'$ data are available in \citet{Cooke2015}. The $J$ band image was obtained with the long-slit intermediate resolution infrared spectrograph (LIRIS), with an exposure time of 8160\,s, and reduced in the standard way using the publicly available program THELI \citep{Erben2005,Schirmer2013}. A $3\sigma$ depth is reached at $i'=26.0$\,mag and $J=23.6$\,mag, with seeing of $\sim 0.76$\,arcsec for both images. 

The IRAC images have a much broader point-spread function (PSF) than the $i'$ and $J$ images, so selecting sources using the 4.5\,\micron image is prone to blending and some galaxies may be missed which are distinct in the $i'$ or $J$ bands. Using solely the $i'$ band to detect sources would result in biasing the selection towards intrinsically blue sources, whereas the $J$ image is relatively shallow, so we use a deep F140W image as a detection image. 
The field around \7C was imaged with the F140W filter of the \emph{Hubble Space Telescope} Wide-Field Camera 3 (\emph{HST}/WFC3) in July 2015  as part of an on-going 40-orbit spectroscopic program (P.I. D. Stern). The \emph{HST} spectra and photometry will be discussed in a future paper (Noirot et al., in preparation). We retrieved the calibrated, dither-combined (drizzled) image from MAST\footnote{Mikulski Archive for Space Telescopes: https://archive.stsci.edu} to use as a detection image. This image has 0.5\,ksec exposure and is complete\footnote{The histogram of number counts per magnitude bin starts to decrease after 25\,mag in F140W.} to at least 24\,mag. 

Source extraction was done with SExtractor \citep{Bertin1996} in dual-image mode. The \emph{HST} F140W detection image was used to detect sources, and photometry was obtained on the $i'$, $J$, 3.6\,\micron and 4.5\,\micron images in 2\,arcsec diameter apertures. 
We are unable to use large apertures to measure the IRAC fluxes (e.g. 4\,arcsec) due to the high spatial density of sources in the cluster core (galaxies are typically $\sim2$\,arcsec apart). So we measure the fluxes in 2\,arcsec diameter apertures and correct for the broader PSF of the IRAC data compared to the ground-based data using the ratio of the flux in the $J$ band image to the $J$ image convolved with a Gaussian kernel matching the IRAC PSF, following \citet{Hartley2013}. Fluxes were corrected to total fluxes using the growth curves of bright, unsaturated stars in the field-of-view.  This method assumes that the blended sources have the same $J-\rm{IRAC}$ colour, so may provide inaccurate colours for some sources. 

\subsection{High redshift galaxy selection}
To select sources likely to lie at high redshift, we employ two colour cuts. The well-tested IRAC cut of $[3.6]-[4.5]>-0.1$ \citep{Papovich2008} selects sources at $z>1.3$ due to the 1.6\,\micron peak of stellar emission moving into the IRAC bands at these redshifts. This cut was adjusted to $[3.6]-[4.5]>-0.2$ in order to be sure of selecting as complete a cluster sample as possible, although this also allows more lower-redshift sources to contaminate the sample. Most stars have $[3.6]-[4.5] \sim -0.5$ and so will be removed by this cut \citep{Galametz2012}. A second cut of $i'-[3.6] > -0.5\times[3.6]+11.4$ \citep{Cooke2015} was applied to further remove bright low-redshift interlopers and contaminating AGN. 
To remove faint sources with potentially inaccurate flux measurements, only those with magnitudes brighter than $23.8$\,mag in $4.5$\,\micron ($5\sigma$ image depth) and $23.6$\,mag in $J$ ($3\sigma$ image depth) were considered. Any sources referred to hereafter are those that match these criteria. 
To maximise the overdensity of (proto)cluster sources to field contaminants, we only consider sources within 0.9\,arcmin of the central RLAGN. 

The completeness of our catalogue is a function of F140W magnitude. 
This image is deeper than the ground-based imaging and is 100\% complete to the $J$ band limit of $23.6$\,mag.

\subsection{Statistical subtraction}
\subsubsection{Control field}
Throughout this study, we use the eighth data release (DR8) of the UKIDSS Ultra Deep Survey \citep[UDS;][]{Hartley2013} as a control field. The IRAC fluxes in the UDS catalogue were deblended using the same method described above, using the resolved $K_s$ images to deblend sources \citep[see][]{Hartley2013}. The UDS has $5\sigma$ depths of $i'=27.0$\,mag, $J=24.9$\,mag, $[3.6]=24.2$\,mag and $[4.5]=24.0$\,mag \citep{Furusawa2008,Hartley2013}. We therefore use the same magnitude cuts for the UDS as for the \7C field. We do not consider Eddington bias due to the greater depth of the UDS, however we do not expect it to significantly affect our results, particularly regarding the fraction of the most massive galaxies in the cluster that are quiescent. 

The UDS is a $K_s$-selected survey, whereas we use an F140W selection for \7C. 
Both selections are done in the infrared, and both are much deeper than our $J \le 23.6$\,mag selection so these methods are unlikely to differ greatly. They would only differ for extremely red sources with very faint F140W magnitudes and bright $K_s$ magnitudes. 
We have checked that the different selection methods do not affect our use of the UDS as a control sample by comparing number counts in the two fields as a function of $i'$, $J$, $[3.6]$ and $[4.5]$ magnitudes and find that they match well within the colour and magnitude cuts stated above. 

Where we compare the properties of the cluster galaxies to the field, we use only those galaxies selected in the UDS which have photometric redshifts between $1.5<z<1.7$.\footnote{photometric redshifts determined from the full 11-band photometry of the UDS, see \citet{Hartley2013}} This ensures we are comparing the cluster properties to those of the field at approximately the same redshift. For statistical subtraction (see below), we use the full UDS with no photometric redshift constraints. \\

\subsubsection{Subtraction of field contaminants}
Since we do not identify cluster members with spectra or photometric redshifts, we use statistical subtraction to derive the cluster galaxy properties. The fore- and background population is estimated from $\sim400$ random 0.9\,arcmin radius regions in the UDS, having applied the same colour and magnitude cuts as above. 
The field contribution is estimated from the median of these 400 regions and then subtracted from the corresponding number of galaxies around \7C. The uncertainty is the 1$\sigma$ standard deviation of the 400 field regions.

\section{Redshift of 7C\,1753+6311} \label{sec:spectrum}
\begin{figure}
\centering
\includegraphics[width=\columnwidth]{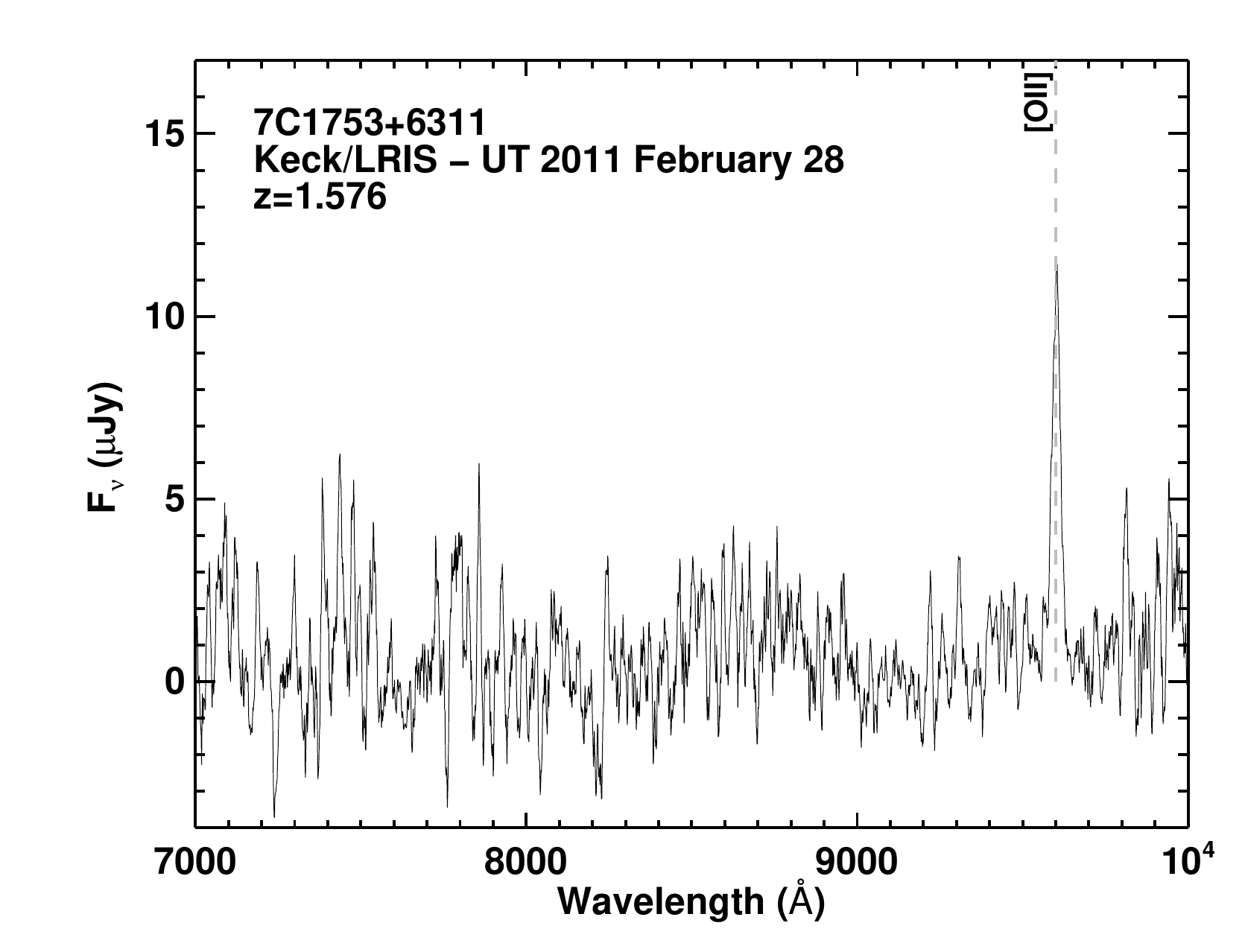}
\caption{The Keck/LRIS spectrum of 7C\,1753+6311 obtained on UT 2011 February 28 (smoothed with a boxcar average of 10\,\AA\ for
clarity). A single, high equivalent width emission line is detected at 9602\,\AA\, which we identify as the [\ion{O}{2}]~$\lambda 3727$ doublet which places 7C\,1753+6311 at $z = 1.576$. This redshift is confirmed with corresponding H$\alpha$ emission by Rettura et al. (in preparation). }\label{fig:spectrum}

\end{figure}

\begin{figure*}
\centering
\includegraphics[trim = 0mm 12.5mm -5mm 0mm, clip, width=1.035\columnwidth]{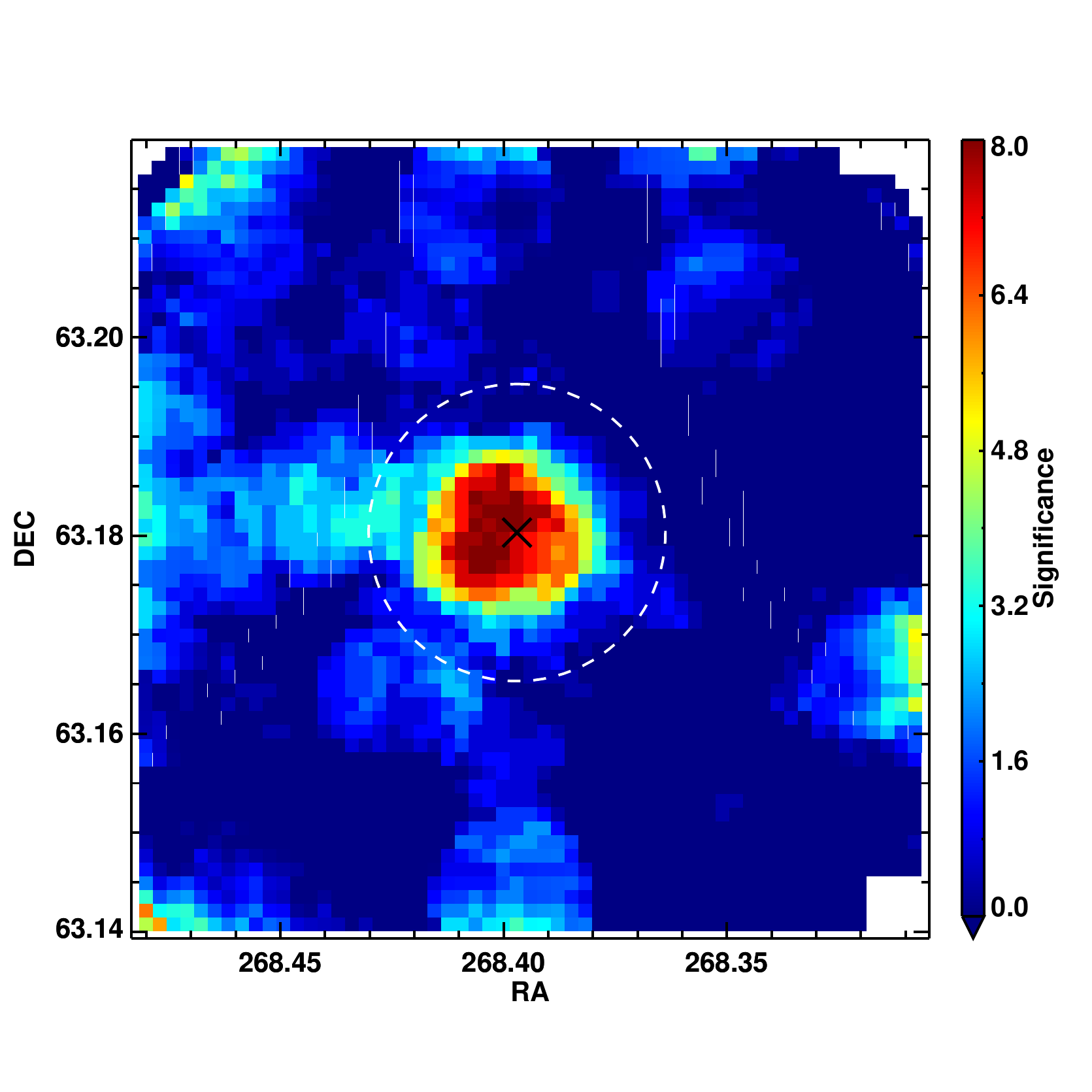}
\includegraphics[trim = 0mm 12.5mm -5mm 0mm, clip, width=1.035\columnwidth]{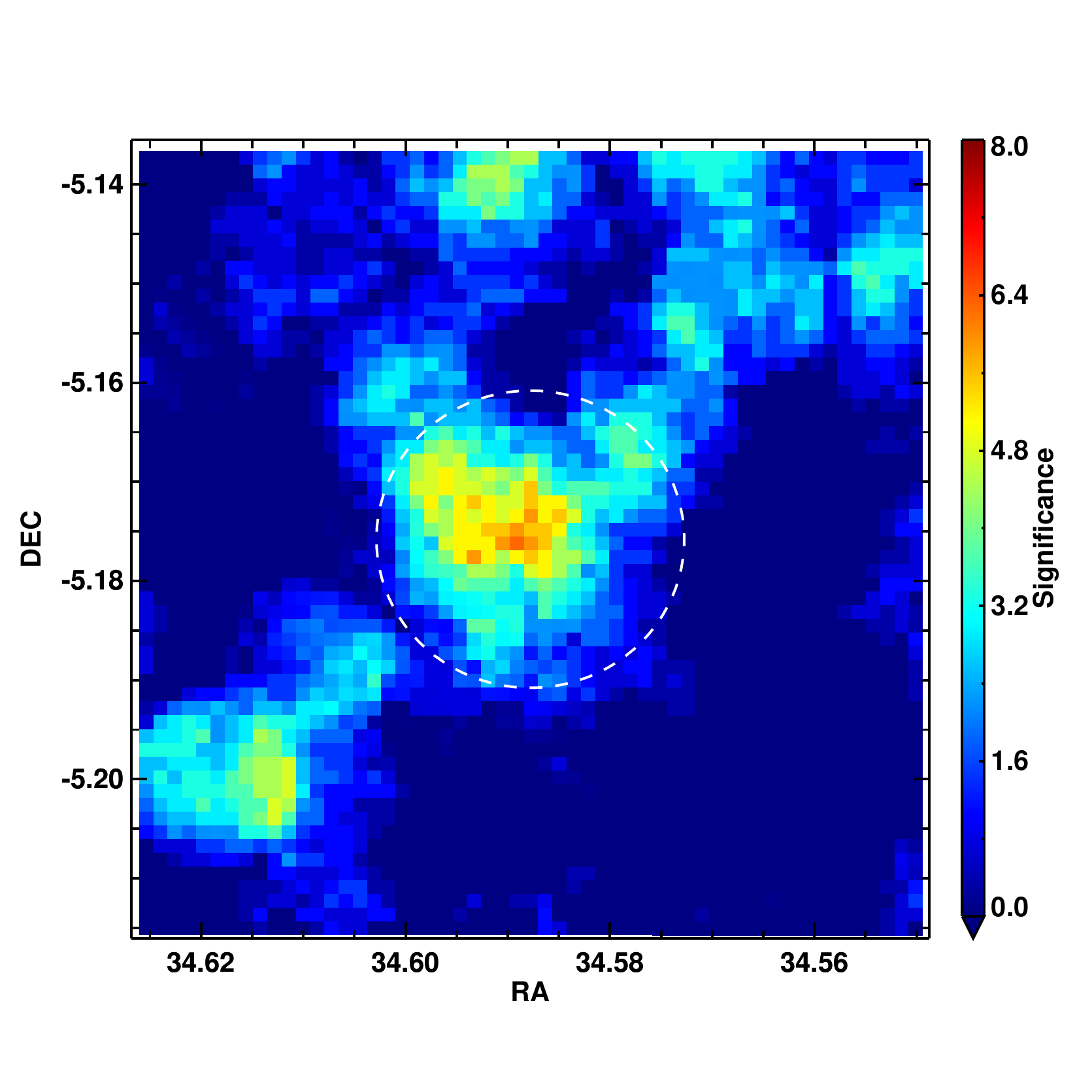}
\caption{{\it{Left}}: Density map of colour-selected sources in \ClG7C. The \emph{HST} image used to detect sources has been supplemented by the $i'$ and $J$ images in the outskirts to show the extended field around \7C. The RLAGN is shown by the black cross and the circle shows a 0.9\,arcmin radius around the RLAGN. Only sources within this circle are considered in this paper. The colourbar shows the overdensity in sigma compared to the average field. {\it{Right}}: Density map of colour-selected sources in ClG\,0218.3$-$0510. The white circle has a radius of 0.9\,arcmin around centre of the cluster, as measured by \citet{Papovich2010}.}
\label{fig:density}
\end{figure*}

We obtained a deep optical spectrum of \7C using the Low Resolution Imaging Spectrometer (LRIS; Oke et al. 1995) at the Keck~I telescope during twilight on UT 2011 February 28.  LRIS is a double spectrograph, and we integrated for 1200\,s on the blue arm and 1120\,s on the red arm in order to match read-out times.  The observations used the 1.5\arcsec\ wide longslit and the data were processed using standard procedures and flux calibrated using an archival sensitivity function from 2011 April.  Nothing is detected on the blue side, but a single, strong, high equivalent width emission line is detected at 9602\,\AA\ on the red side, which we identify with the [\ion{O}{2}]~$\lambda 3727$ doublet at $z = 1.576$ (see Figure\,\ref{fig:spectrum}). For the instrument configuration we used, the spectral resolving power was
$R \equiv \lambda / \Delta \lambda = 1600$ for objects filling the slit at $\sim 9600$\,\AA, which is insufficient to resolve the [\ion{O}{2}] doublet.  However, the redshift is confirmed by the detection of corresponding H$\alpha$ emission in a Keck near-infrared spectrum reported by Rettura et al. (in preparation). 
In comparison to radio galaxies surrounded by protoclusters typically reported in the literature \citep[e.g.][]{Venemans2007,Galametz2010a,Hatch2011a}, this is a relatively weak line emitter. 
We do not find any features at 4854\,\AA\ and this wavelength does not correspond to any strong spectral features for our measured redshift. The feature noted in \citet{Lacy1999} was therefore probably due to noise in their shallow data. 

\section{Cluster properties} \label{sec:results}
\subsection{A galaxy cluster at $z=1.58$} 
Figure \ref{fig:density} shows a density map of the field around \7C and ClG\,0218.3$-$0510, a well-studied cluster at $z=1.62$ \citep{Papovich2010,Tanaka2010}. These maps were produced by measuring the number density of sources (selected using the colour and magnitude criteria described in Section \ref{sec:method}) within 30\,arcsec radius apertures around each 5\,arcsec pixel. The UDS was mapped in the same way, and the mean and standard deviation of densities in the UDS were used to convert each number density value to a significance above the expected field density. The pixels in Figure \ref{fig:density} are therefore correlated as each 5\,arcsec pixel indicates the overdensity within a 30\,arcsec radius aperture. Using the selection criteria defined in Section \ref{sec:method}, the peak overdensity around \7C is an $8.9\sigma$ significance of galaxies within a 30\,arcsec aperture, centred 16\,arcsec (136\,kpc) from the radio galaxy. 
These galaxies appear to be highly clustered around the central RLAGN (Figure \ref{fig:fieldimg}). The source density is so high in the central 0.9\,arcmin region that five pairs of sources were blended in the 4.5\,\micron image, and their true nature was only discovered in the higher resolution ground-based data.  It is possible that many of the other CARLA cluster candidates which have extremely high galaxy overdensities also suffer from blending.

Follow-up near-infrared Keck spectroscopy of \7C revealed five galaxies, including the RLAGN, with spectroscopic redshifts between $1.578<z<1.587$ within a projected diameter of 2\,Mpc (Rettura et al., in preparation). This structure therefore satisfies the criteria set out by \citet{Eisenhardt2008} for a spectroscopically confirmed $z>1$ (proto)cluster, and so we refer to the structure as \ClG7C from now on. 

Besides the RLAGN, there are $29\pm 6$ excess galaxies within 0.9\,arcmin of \7C that are selected with the above colour and magnitude criteria\footnote{The number of excess galaxies was calculated by taking the number of field galaxies selected in 400 random 0.9\,arcmin fields in the UDS and subtracting this from the number of galaxies selected around \7C, then taking the mean and standard deviation of the resultant distribution.}. 
This level of clustering and overdensity is slightly greater than that of the ClG\,0218.3$-$0530 protocluster at $z=1.62$
, which has a galaxy excess of $22\pm 6$ using the same criteria. The cluster ClG\,0218.3$-$0530  is a well-studied structure with a tentative 4.5$\sigma$ X-ray detection potentially indicating a collapsed core \citep{Tanaka2010}. 
The comparably high galaxy overdensity surrounding \7C (Figure \ref{fig:density}) suggests that this RLAGN is surrounded by a protocluster consisting of a dominant main halo that is already a relatively high-mass group. 

The approximate mass of \ClG7C can be estimated from the galaxy richness. \citet{Andreon2014b} reported that galaxy richness was a good proxy for cluster mass, with little dependence on redshift. ClG\,0218.3$-$0530 has an X-ray determined mass of 4-8$\times10^{13}$\,\Msun\ \citep{Tanaka2010,Pierre2012}, which is consistent with the galaxy velocity dispersion \citep{Tran2015}. Since \ClG7C is richer than ClG\,0218.3$-$0530, its mass is likely to be slightly greater. Using equation 3 from \citet{Andreon2014b}, and the calculated value of $m^*_{4.5\mu m}+1 = 21.2$ from  \citet{Wylezalek2014} we measure a cluster richness of $10\pm1$ galaxies\footnote{This is the number of background-subtracted galaxies in the \7C field.} with $[4.5]\le21.2$, and estimate the mass within 500kpc of \7C to be $(9.2\pm4.5)\times10^{13}$\,M$_{\odot}$, consistent with this structure being a slightly more massive group than ClG\,0218.3$-$0530.

\subsection{Red sequence and red fraction} \label{sec:CMD} 
One of the signs of a mature cluster is the presence of a red sequence. %
Ubiquitous in clusters at $z<1$, red sequences persist in galaxy clusters out to at least $z=1.4$ \citep{Stanford2005,Snyder2012}, and have been found in some dense (proto)clusters at even higher redshifts \citep[e.g.][]{Kodama2007,Stanford2012,Newman2014}. 

\begin{figure}
\centering
\includegraphics[width=\columnwidth]{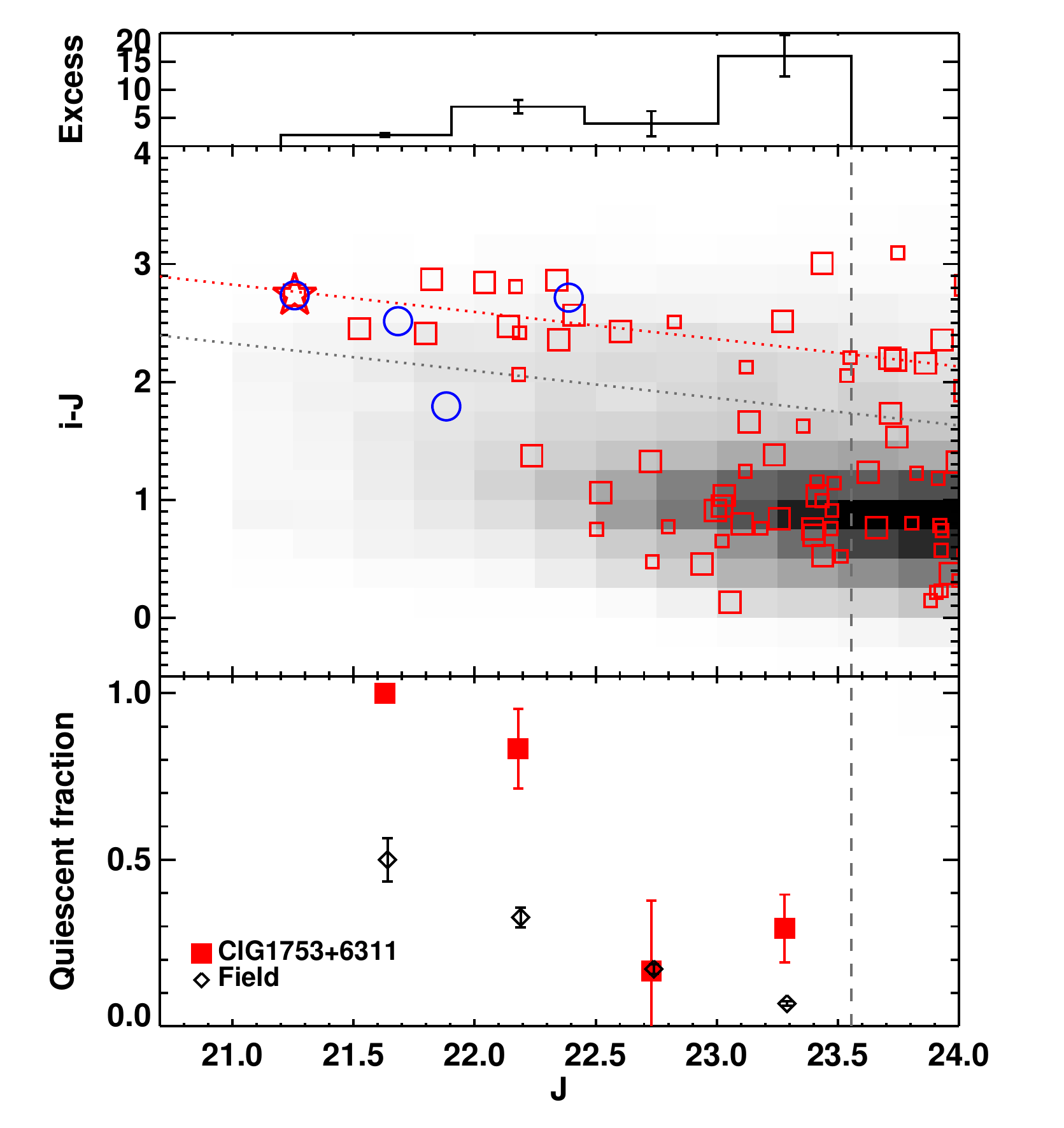}
\caption{{\it{Top}}: Histogram showing the number of excess sources within 0.9\,arcmin of \7C compared to the expected numbers in a random blank field, as a function of $J$ band magnitude. 
{\it{Middle}}: Colour-magnitude diagram showing a clear structure of red sources at $i'-J \sim 2.5$. The RLAGN is marked with a red star. All sources within 0.9\,arcmin of the RLAGN are marked with red squares. Larger squares indicate those within 0.5\,arcmin of the RLAGN. The grey dashed line indicates the $3\sigma$ depth of the $J$ band. The best fit to the red sequence is shown by the red dotted line. The grey dotted line indicates 0.5\,mag below this line. The background greyscale shows the normalized distribution of the UDS for comparison. 
Spectroscopic members (Rettura et al., in preparation) are highlighted with large blue circles. 
{\it{Bottom}}: The quiescent fraction of galaxies in \ClG7C as a function of $J$ band magnitude. Red squares show the cluster values, black diamonds indicate the quiescent fractions measured for field galaxies. 
\vspace{1mm}}
\label{fig:CMD}
\end{figure}

In Figure \ref{fig:CMD} we show the $i'-J$ colour-magnitude diagram of sources within 0.9\,arcmin of the RLAGN ($\sim500$\,kpc at this redshift)\footnote{None of our results qualitatively change when we consider a smaller 30\,arcsec radius field.}. Larger squares indicate sources that are within 30\,arcsec of the radio galaxy. The histogram in the top panel of Figure \ref{fig:CMD} shows the excess number of galaxies in \ClG7C, compared to the UDS control field. There is a significant overdensity in the field around \7C at all magnitudes, increasing at the faint end. Although we expect contamination from fore- and background field sources in the colour-magnitude diagram, the majority of the red data points are likely to be cluster members, and there is a clear, strong red sequence at $J<23$\,mag, with hints of the sequence continuing to fainter magnitudes.

The red sequence is fit by the line $i'-J = 7.688 - 0.232 \times J$, calculated by iteratively clipping sources more than 1.5$\sigma$ from the best-fit line, allowing both the slope and normalization to freely vary, until convergence was reached. The fit is 
shown by the red dotted line in Figure \ref{fig:CMD}. The colours of sources on this red sequence suggest they formed at redshifts of $2<z_f<3$, assuming the galaxies formed their stars in single bursts. A cluster formation model in which the member galaxies formed over the course of 2-3\,Gyr, with galaxy formation peaking at $z=3$ predicts an average red sequence colour of $i'-J=2.7$\,mag, consistent with the data \citep{Cooke2015}. 

We define ``red" galaxies as those that lie within 0.5\,mag of (or redder than) the red sequence (shown by the lower, grey, dotted line in Figure \ref{fig:CMD}); this cleanly divides the red sequence from the blue cloud. 
We calculate the red fraction for \ClG7C, statistically removing the expected number of field contaminants as: 
\begin{equation}
f_{\rm red} = \langle\frac{(N^{\rm 7C1753}_{\rm red} - N^{\rm field}_{\rm red})}{(N^{\rm 7C1753}_{\rm total} - N^{\rm field}_{\rm total})}\rangle_{\rm median}
\end{equation}

where $N^{\rm field}_{\rm red}$ and $N^{\rm field}_{\rm total}$ are the measured number of ``red"  and total galaxies that satisfy our colour criteria in $\sim400$ random 0.9\,arcmin field regions from the UDS. The uncertainty is the $1\sigma$ standard deviation in the calculated red fractions for \ClG7C. 

The fraction of red galaxies in the protocluster is significantly larger than in the blank field. The red fraction of \ClG7C, after statistically removing field contaminants, is $f_{\rm red} = 0.66 \pm 0.13$, compared to the average fraction of $1.5<z<1.7$ galaxies in the UDS control field, which is $f_{\rm red} = 0.27 \pm 0.01$ (see Table \ref{table:passivefraction}). 
\ClG7C has a similar red fraction to the $z=1.62$ protocluster ClG\,0218.3$-$0510, which has $f_{\rm red} = 0.48 \pm 0.15$. So the enhanced red fraction in \ClG7C seems typical for mature protoclusters. The dense environment of the protoclusters appears to have a strong impact on the colours of their member galaxies.

\subsection{Quiescent galaxy fraction}\label{sec:UVJ}

In low-redshift clusters the galaxies that lie on the red sequence are predominantly passively evolving, old galaxies. However, dusty star forming galaxies (with $A_V\sim1$-3) exhibit colours similar to those expected from quenched, passively evolving (i.e. quiescent) galaxies, and these galaxies make up approximately half of the red infrared-selected galaxy population at higher redshifts \citep{Kriek2008}.
Furthermore, recent literature has also shown that high-redshift clusters and protoclusters do contain dusty star-forming galaxies \citep[e.g.][]{Brodwin2013,Dannerbauer2014,Smail2014,Santos2015}. The enhanced red fraction of sources in \ClG7C could therefore be ascribed to an excess of dusty star-forming galaxies and/or quenched, passively evolving galaxies. 
Here we use the rest-frame $U,B,J$ colours (observed $i',J,[3.6]$) to separate these two populations.

Using the method outlined in \citet{Williams2009a}, \citet{Papovich2012} used the observed-frame $z'$, $J$ and 3.6\,\micron bands (rest-frame $U,B,J$) to separate galaxies in the $z=1.62$ cluster ClG\,0218.3$-$0510 into quiescent and star forming populations. Using the full spectral energy distribution (SED) fits to the ClG\,0218.3$-$0510 cluster members, we have converted the \citet{Papovich2012} selection criteria to use our $i'$, $J$ and 3.6\,\micron bands. 
Quiescent galaxies are those which satisfy the following (slightly stricter) criteria:

\begin{equation}
i'-J \ge 2.0 
\end{equation}
\begin{equation}
J-[3.6] \le 1.7 
\end{equation}
\begin{equation}
i'-J \ge 0.375+1.25 \times (J-[3.6])
\end{equation}

We caution the reader that these equations were derived specifically for the eighth data release of the UDS and our data. The 3.6\,\micron magnitudes may be systematically offset by up to 0.5\,mag due to the method by which they were determined and so these criteria may change for different datasets.

\begin{figure}
\centering
\includegraphics[width=\columnwidth]{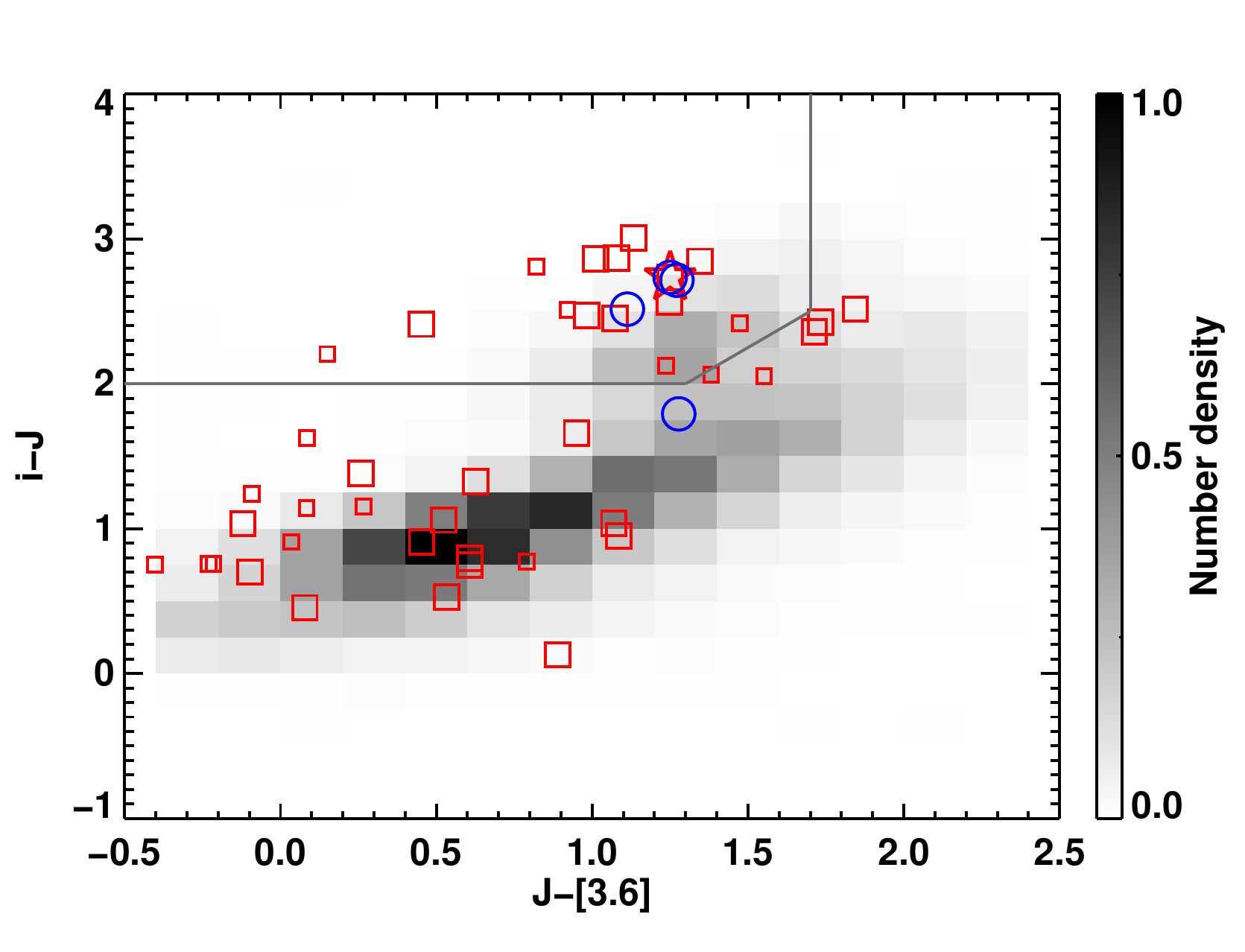}
\caption{Observed $i'-J$ versus $J-[3.6]$ (rest-frame $U-B$ versus $B-J$) colour-colour diagram. The upper left quadrant selects quiescent galaxies at $z\sim1.6$. Sources further towards the upper-right are dusty star forming objects and those at bluer $i'-J$ colours are star forming galaxies. The background density map shows the expected normalized distribution of field sources from the UDS. The symbols are the same as in Figure \ref{fig:CMD}.}
\label{fig:UVJ}
\end{figure}

Figure \ref{fig:UVJ} shows the distribution of all sources selected in \ClG7C in $i'-J$ versus $J-[3.6]$ (rest-frame $U-B$ versus $B-J$) colour-colour space. The greyscale shows the expected distribution of the control field. The lines show the $i',J,[3.6]$ criteria used to select quiescent galaxies, which lie in the upper-left quadrant. 
The full cluster membership of \ClG7C is not known, so interlopers were statistically removed in $i'J[3.6]$ colour-colour space using the UDS as the control field. 
To do this we use $\sim400$ random 0.9\,arcmin radius regions in the UDS, classifying sources as ``quiescent" or ``star forming" using the above criteria. 
Sources in the \7C field are then classified as ``quiescent" or ``star forming", and the quiescent fraction calculated as:
\begin{equation}
f_{\rm Q} = \langle\frac{(N^{\rm 7C1753}_{\rm Q} - N^{\rm field}_{\rm Q})}{(N^{\rm 7C1753}_{\rm total} - N^{\rm field}_{\rm total})}\rangle_{\rm median}
\end{equation}
where $N^{\rm field}_{\rm Q}$ is the measured number of rest-frame $UBJ$-selected quiescent galaxies in $\sim400$ random 0.9\,arcmin field regions. The uncertainty is the $1\sigma$ standard deviation in the calculated quiescent fractions for \ClG7C. 

Without far-IR data, we are unable to locate extremely dust-obscured systems, so these will not be found in either the \7C field or UDS and would be missing from Figure \ref{fig:UVJ}. 
These extremely dusty galaxies are rare, but could be an important population in protoclusters \citep[e.g.][]{Brodwin2013}. In addition, some galaxies (of order $\sim10\%$) may be misclassified due to very dusty regions within them. Further analysis with submillimeter data 
would be required to examine the extremely dusty populations in these fields. This means that we cannot analyse the extremely dust-obscured populations in any of the fields considered here, but we are able to do a robust comparison between them as the dusty populations are undetected in all of these fields: \ClG7C, ClG\,0218.3$-$0510 and the UDS control field. 

Half of the detected galaxies in \ClG7C are quiescent, with a quiescent fraction ($f_Q$) for sources with $J \le 23.6$ of $f_Q = 0.50 \pm 0.09$ (see Table \ref{table:passivefraction}). Of the ``red" galaxies, $80\pm 6$\% are quiescent,  %
so the vast majority of these objects are not dust-obscured star forming galaxies, but are already quenched and evolving passively.  

ClG\,0218.3$-$0510 contains fewer passively evolving galaxies ($f_Q=0.30\pm0.08$), with  $67\pm11$\% of the red galaxies classified as quiescent. These fractions were calculated using the same criteria as in Section \ref{sec:method} and within a 0.9\,arcmin aperture of the cluster core. \ClG7C has a similar fraction of red, quiescent galaxies at a 1$\sigma$ level. 

Both of these protoclusters contain a significantly higher quiescent fraction than the average field, which is $f_Q=0.16\pm0.01$. 
This means that the star formation rates of many cluster members are greatly suppressed relative to the field. 

\begin{figure}
\centering
\includegraphics[width=\columnwidth]{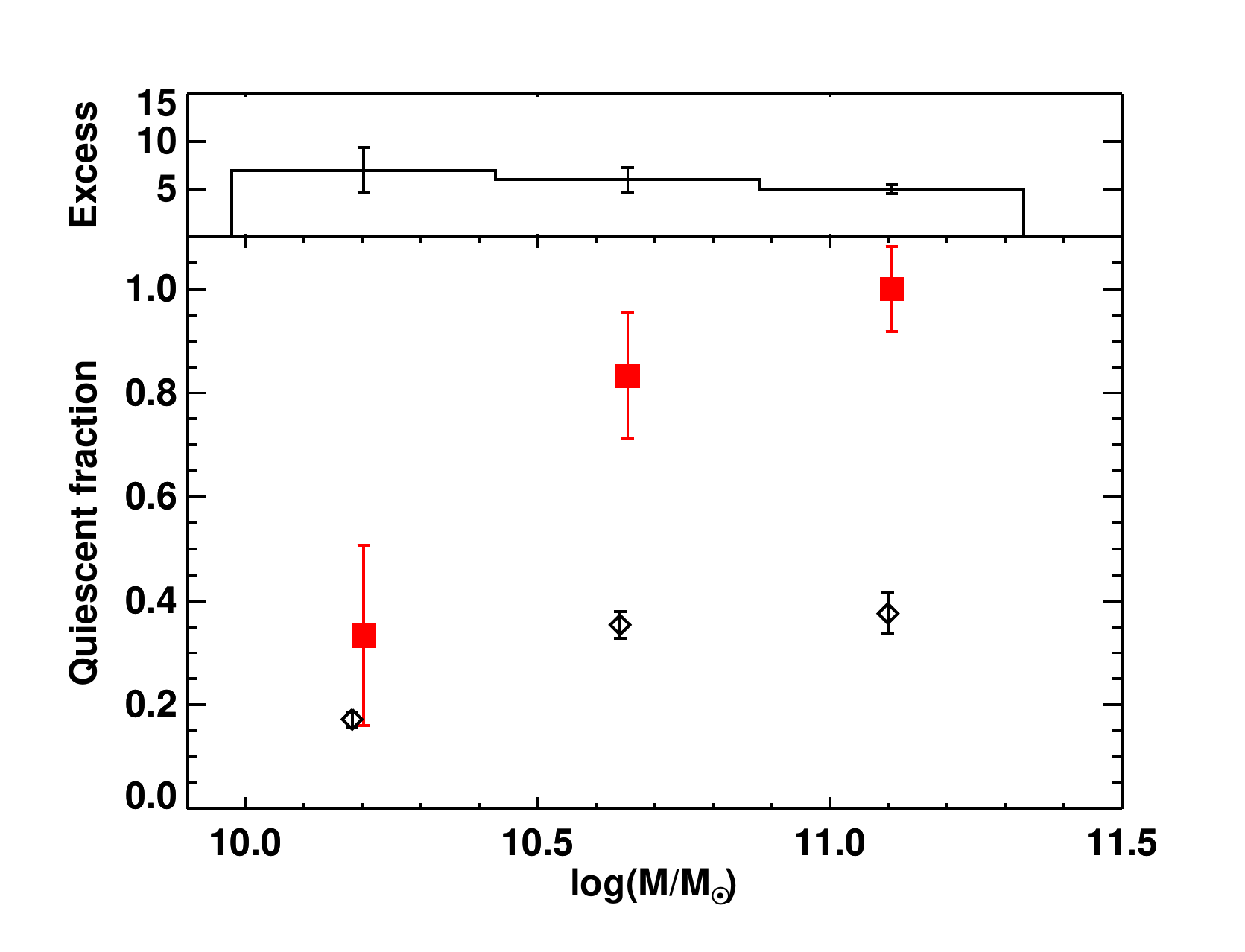}
\caption{The quiescent fraction measured in 4.5\,\micron magnitude bins, which correspond to stellar mass bins. Red squares show the cluster values, black diamonds indicate the quiescent fractions measured for field galaxies. 
The top histogram shows the excess number of sources in \ClG7C, compared to a random blank field, per mass bin.}
\label{fig:passivefraction}
\end{figure}

\subsection{Passive fraction as a function of mass}
The quiescent fraction is a strong function of $J$ band magnitude. As shown in the bottom panel of Figure \ref{fig:CMD}, the quiescent fraction gradually rises with decreasing magnitude. %
At $J<22.5$\,mag the fraction of quiescent galaxies in the protocluster rises to $>80$\%, double the field fraction. 

The $[4.5]$ flux provides a better correlation with stellar mass than the $J$ flux, and is nearly independent of galaxy type. Using galaxies with known stellar masses (from full SED-fitting) in the UDS \citep{Mortlock2013}, we convert the [4.5] magnitudes to stellar mass using $\log ( M_* / {\rm M}_{\odot} )= 22.53-0.57\times[4.5]$. This equation was derived empirically by fitting a line to the stellar masses and [4.5] magnitudes from the UDS, which was resampled to have the same $J\le 23.6$ quiescent fraction as \ClG7C, i.e. $50\%$, to remove the slight dependence of this relation on galaxy type. This line has an intrinsic scatter of $0.2$\,dex. We use this equation to calculate the masses for \ClG7C. We use a similar equation but for the full UDS with no resampling, to calculate the masses for the field. This simply corresponds to a slight shift in the bin centre in Figure \ref{fig:passivefraction}. 
Using these equations, we recalculate the quiescent fractions as a function of stellar mass (Figure \ref{fig:passivefraction}). These fractions were calculated as in Section \ref{sec:UVJ} for sources in [4.5] magnitude bins (corresponding to stellar mass). 

There is also a strong correlation between galaxy mass and passivity, with a higher fraction of the massive sources being quiescent.  Figure \ref{fig:passivefraction} shows that this trend is steeper for the protocluster galaxies than for the field galaxies, and there is a divide at $M_* \sim 10^{10.5}$\,M$_{\odot}$. Only $20-30$\% of galaxies with stellar masses $M_* < 10^{10.5}$\,M$_{\odot}$ are quiescent in both environments, whereas 80-100\% of  $M_* > 10^{10.5}$\,M$_{\odot}$ galaxies are quiescent in the protocluster, compared to only $\sim40$\% in the field (Table \ref{table:passivefraction}).

We find that the fraction of quiescent galaxies is dependent on environment: \ClG7C contains double the quiescent fraction of the control field at $z\sim1.6$. However, this environmental effect is also mass dependent: only the population of high-mass galaxies has an enhanced quenched fraction relative to the control field. 
These results are consistent with recent literature on red galaxies in clusters at $z>1.5$. \citet{Rudnick2012} and \citet{Fassbender2014} found a strong excess of bright, red galaxies in two $z\sim1.6$ clusters, but a corresponding lack of faint, red galaxies. However, in contrast to these results, \citet{Andreon2014a} found a well-populated red sequence down to $\sim10^{10}$\,\Msun\ in a $z\sim1.8$ cluster and \citet{Lee2015} find that there is no difference in the quiescent fraction between cluster and field environments at $z>1$, with a large variation between individual clusters. Therefore the mass dependence of quiescent galaxies needs to be analysed in a larger sample of protoclusters to draw firm conclusions. 

In \citet{VanderBurg2013} clusters at $z\sim1$ were also shown to have an increased quiescent fraction compared to the field. The quiescent fractions in \ClG7C are similar to the $z\sim1$ fractions at high masses ($M>10^{10.5}$\,\Msun), which is further evidence that \ClG7C is already a very mature structure, more similar to $z=1$ clusters than higher redshift protoclusters. The quiescent fraction at lower masses is much higher at $z\sim1$ than in \ClG7C. This may suggest a build up of the low mass end of the red sequence in clusters from $z=1.6$ to $z=1$. 

\begin{table}
\begin{centering}

\caption{Fractions of quiescent galaxies in \ClG7C, ClG\,0218.3$-$0510 and in the control field UDS at $1.5<z<1.7$. $f_{\rm Q}$ gives the quiescent galaxy fraction and $f_{\rm red}$ gives the fraction of galaxies with red colours in each sample.\\
}
\label{table:passivefraction}
\begin{tabular}{ l c c c c }
\hline
 & \hspace{-0.5cm}CARLA\,J1753 & ClG0218 & UDS \\ \hline \hline
$f_{\rm red}$ & \hspace{-0.5cm}$0.66 \pm 0.13$ & $0.48 \pm 0.15$ & $0.27 \pm 0.01$  \\
$f_{\rm Q}$ ($J \le 23.6$) & \hspace{-0.5cm}$0.50 \pm 0.09$ & $0.30 \pm 0.08$ & $0.16 \pm 0.01$  \\
$f_{\rm Q}$ (red galaxies)& \hspace{-0.5cm}$0.80 \pm 0.06$ & $0.67 \pm 0.11$ & $0.61 \pm 0.03$ \\
$f_{\rm Q}$ ($M_* \ge 10^{10}$\Msun)& \hspace{-0.5cm}$0.76 \pm 0.13$ & $0.44 \pm 0.14$ & $0.28 \pm 0.01$ \\
$f_{\rm Q}$ ($M_* \ge 10^{10.5}$\Msun)& \hspace{-0.5cm}$0.91 \pm 0.09$ & $0.38 \pm 0.16$ & $0.36 \pm 0.02$ \\
\end{tabular}
\end{centering}
\end{table}

\section{Conclusions} \label{sec:conclusion}
We present the first robust spectroscopic redshift of the high redshift RLAGN \7C, placing it at $z=1.58$. We show this radio galaxy is located in an $8.9\sigma$ galaxy overdensity, implying that it is embedded in a high redshift galaxy cluster. 
The cluster core contains $28\pm6$ excess galaxies brighter than $J=23.6$\,mag. This galaxy richness implies a cluster mass of at least several $\times10^{13}$\,M$_{\odot}$. Of these excess galaxies, $66\pm13$\% have red colours and lie on a sequence in colour-magnitude space. %
The rest-frame $UBJ$ colours of these galaxies show that $80$\% of the red galaxies are quiescent, therefore this is a mature cluster with a predominantly old stellar population. More than 80\% of the galaxies with masses $M_* > 10^{10.5}$\,M$_{\odot}$  are quiescent in this cluster, compared to only $\sim40\%$ of field galaxies of this high mass. At lower masses we find no difference between the quiescent fractions of the field and cluster galaxies. This mature structure is similar in the level of clustering, overdensity and red fraction to other clusters at a similar redshift. %
The presence of a dense core and a well-formed, passively evolving red sequence suggest that RLAGN do not solely reside in young, uncollapsed protoclusters, rather they can be used as beacons for clusters in a wide range of evolutionary states. 

\acknowledgements

The authors would like to thank Omar Almaini for making the UDS catalogues and images available and the staff at the WHT for taking the service mode observations. 
We thank the anonymous referee for their careful review of the manuscript and their helpful comments which improved the content of the paper. 

EAC acknowledges the support of the STFC. NAH is supported by an STFC Rutherford Fellowship. 
The work of DS was carried out at the Jet Propulsion Laboratory, California Institute of Technology, under a contract with NASA. 

This work was based on observations made with the William Herschel Telescope under programme IDs W/2013b/10 and SW2015a07 and with the \emph{Spitzer Space Telescope}. The William Herschel Telescope and its service programme are operated on the island of La Palma by the Isaac Newton Group in the Spanish Observatorio del Roque de los Muchachos of the Instituto de Astrof\'{i}sica de Canarias. \emph{Spitzer} is operated by the Jet Propulsion Laboratory, California Institute of Technology under a contract with NASA. Support for this work was provided by NASA through an award issued by JPL/Caltech. 

Some of the data presented herein were obtained at the W.M. Keck Observatory, which is operated as a scientific partnership among the California Institute of Technology, the University of California and the National Aeronautics and Space Administration. The Observatory was made possible by the generous financial support of the W.M. Keck Foundation.  The authors wish to recognize and acknowledge the very significant cultural role and reverence that the summit of Mauna Kea has always had within the indigenous Hawaiian community.  We are very grateful to have the opportunity to conduct observations from this mountain.

\bibliographystyle{mn2e}
\bibliography{protoclustersbib}

\begin{thebibliography}{58}
\expandafter\ifx\csname natexlab\endcsname\relax\def\natexlab#1{#1}\fi

\bibitem[{{Andreon} \& {Congdon}(2014)}]{Andreon2014b}
{Andreon} S., {Congdon} P., 2014, \aap, 568, A23

\bibitem[{{Andreon} {et~al}\mbox{.}(2014){Andreon}, {Newman}, {Trinchieri},
  {Raichoor}, {Ellis}, \& {Treu}}]{Andreon2014a}
{Andreon} S., {Newman} A.~B., {Trinchieri} G., {Raichoor} A., {Ellis} R.~S.,
  {Treu} T., 2014, \aap, 565, A120

\bibitem[{{Bertin} \& {Arnouts}(1996)}]{Bertin1996}
{Bertin} E., {Arnouts} S., 1996, \aaps, 117, 393

\bibitem[{{Bleem} {et~al}\mbox{.}(2015){Bleem}, {Stalder}, {de Haan}, {Aird},
  {Allen}, {Applegate}, {Ashby}, {Bautz}, {Bayliss}, {Benson}, {Bocquet},
  {Brodwin}, {Carlstrom}, {Chang}, {Chiu}, {Cho}, {Clocchiatti}, {Crawford},
  {Crites}, {Desai}, {Dietrich}, {Dobbs}, {Foley}, {Forman}, {George},
  {Gladders}, {Gonzalez}, {Halverson}, {Hennig}, {Hoekstra}, {Holder},
  {Holzapfel}, {Hrubes}, {Jones}, {Keisler}, {Knox}, {Lee}, {Leitch}, {Liu},
  {Lueker}, {Luong-Van}, {Mantz}, {Marrone}, {McDonald}, {McMahon}, {Meyer},
  {Mocanu}, {Mohr}, {Murray}, {Padin}, {Pryke}, {Reichardt}, {Rest}, {Ruel},
  {Ruhl}, {Saliwanchik}, {Saro}, {Sayre}, {Schaffer}, {Schrabback},
  {Shirokoff}, {Song}, {Spieler}, {Stanford}, {Staniszewski}, {Stark}, {Story},
  {Stubbs}, {Vanderlinde}, {Vieira}, {Vikhlinin}, {Williamson}, {Zahn}, \&
  {Zenteno}}]{Bleem2015}
{Bleem} L.~E. {et~al.}, 2015, \apjs, 216, 27

\bibitem[{{Brodwin} {et~al}\mbox{.}(2012){Brodwin}, {Gonzalez}, {Stanford},
  {Plagge}, {Marrone}, {Carlstrom}, {Dey}, {Eisenhardt}, {Fedeli}, {Gettings},
  {Jannuzi}, {Joy}, {Leitch}, {Mancone}, {Snyder}, {Stern}, \&
  {Zeimann}}]{Brodwin2012}
{Brodwin} M. {et~al.}, 2012, \apj, 753, 162

\bibitem[{{Brodwin} {et~al}\mbox{.}(2013){Brodwin}, {Stanford}, {Gonzalez},
  {Zeimann}, {Snyder}, {Mancone}, {Pope}, {Eisenhardt}, {Stern}, {Alberts},
  {Ashby}, {Brown}, {Chary}, {Dey}, {Galametz}, {Gettings}, {Jannuzi},
  {Miller}, {Moustakas}, \& {Moustakas}}]{Brodwin2013}
{Brodwin} M. {et~al.}, 2013, \apj, 779, 138

\bibitem[{{Chiang}, {Overzier} \& {Gebhardt}(2013){Chiang}, {Overzier}, \&
  {Gebhardt}}]{Chiang2013}
{Chiang} Y.-K., {Overzier} R., {Gebhardt} K., 2013, \apj, 779, 127

\bibitem[{{Chiang}, {Overzier} \& {Gebhardt}(2014){Chiang}, {Overzier}, \&
  {Gebhardt}}]{Chiang2014}
{Chiang} Y.-K., {Overzier} R., {Gebhardt} K., 2014, \apjl, 782, L3

\bibitem[{{Cooke} {et~al}\mbox{.}(2014){Cooke}, {Hatch}, {Muldrew}, {Rigby}, \&
  {Kurk}}]{Cooke2014}
{Cooke} E.~A., {Hatch} N.~A., {Muldrew} S.~I., {Rigby} E.~E., {Kurk} J.~D.,
  2014, \mnras, 440, 3262

\bibitem[{{Cooke} {et~al}\mbox{.}(2015){Cooke}, {Hatch}, {Rettura},
  {Wylezalek}, {Galametz}, {Stern}, {Brodwin}, {Muldrew}, {Almaini},
  {Conselice}, {Eisenhardt}, {Hartley}, {Jarvis}, {Seymour}, \&
  {Stanford}}]{Cooke2015}
{Cooke} E.~A. {et~al.}, 2015, \mnras, 452, 2318

\bibitem[{{Dannerbauer} {et~al}\mbox{.}(2014){Dannerbauer}, {Kurk}, {De
  Breuck}, {Wylezalek}, {Santos}, {Koyama}, {Seymour}, {Tanaka}, {Hatch},
  {Altieri}, {Coia}, {Galametz}, {Kodama}, {Miley}, {R{\"o}ttgering},
  {Sanchez-Portal}, {Valtchanov}, {Venemans}, \& {Ziegler}}]{Dannerbauer2014}
{Dannerbauer} H. {et~al.}, 2014, \aap, 570, A55

\bibitem[{{Eisenhardt} {et~al}\mbox{.}(2008){Eisenhardt}, {Brodwin},
  {Gonzalez}, {Stanford}, {Stern}, {Barmby}, {Brown}, {Dawson}, {Dey}, {Doi},
  {Galametz}, {Jannuzi}, {Kochanek}, {Meyers}, {Morokuma}, \&
  {Moustakas}}]{Eisenhardt2008}
{Eisenhardt} P.~R.~M. {et~al.}, 2008, \apj, 684, 905

\bibitem[{{Erben} {et~al}\mbox{.}(2005){Erben}, {Schirmer}, {Dietrich},
  {Cordes}, {Haberzettl}, {Hetterscheidt}, {Hildebrandt}, {Schmithuesen},
  {Schneider}, {Simon}, {Deul}, {Hook}, {Kaiser}, {Radovich}, {Benoist},
  {Nonino}, {Olsen}, {Prandoni}, {Wichmann}, {Zaggia}, {Bomans}, {Dettmar}, \&
  {Miralles}}]{Erben2005}
{Erben} T. {et~al.}, 2005, Astronomische Nachrichten, 326, 432

\bibitem[{{Fassbender} {et~al}\mbox{.}(2014){Fassbender}, {Nastasi}, {Santos},
  {Lidman}, {Verdugo}, {Koyama}, {Rosati}, {Pierini}, {Padilla}, {Romeo},
  {Menci}, {Bongiorno}, {Castellano}, {Cerulo}, {Fontana}, {Galametz},
  {Grazian}, {Lamastra}, {Pentericci}, {Sommariva}, {Strazzullo}, {{\v
  S}uhada}, \& {Tozzi}}]{Fassbender2014}
{Fassbender} R. {et~al.}, 2014, \aap, 568, A5

\bibitem[{{Fazio} {et~al}\mbox{.}(2004){Fazio}, {Hora}, {Allen}, {Ashby},
  {Barmby}, {Deutsch}, {Huang}, {Kleiner}, {Marengo}, {Megeath}, {Melnick},
  {Pahre}, {Patten}, {Polizotti}, {Smith}, {Taylor}, {Wang}, {Willner},
  {Hoffmann}, {Pipher}, {Forrest}, {McMurty}, {McCreight}, {McKelvey},
  {McMurray}, {Koch}, {Moseley}, {Arendt}, {Mentzell}, {Marx}, {Losch},
  {Mayman}, {Eichhorn}, {Krebs}, {Jhabvala}, {Gezari}, {Fixsen}, {Flores},
  {Shakoorzadeh}, {Jungo}, {Hakun}, {Workman}, {Karpati}, {Kichak}, {Whitley},
  {Mann}, {Tollestrup}, {Eisenhardt}, {Stern}, {Gorjian}, {Bhattacharya},
  {Carey}, {Nelson}, {Glaccum}, {Lacy}, {Lowrance}, {Laine}, {Reach},
  {Stauffer}, {Surace}, {Wilson}, {Wright}, {Hoffman}, {Domingo}, \&
  {Cohen}}]{Fazio2004}
{Fazio} G.~G. {et~al.}, 2004, \apjs, 154, 10

\bibitem[{{Furusawa} {et~al}\mbox{.}(2008){Furusawa}, {Kosugi}, {Akiyama},
  {Takata}, {Sekiguchi}, {Tanaka}, {Iwata}, {Kajisawa}, {Yasuda}, {Doi},
  {Ouchi}, {Simpson}, {Shimasaku}, {Yamada}, {Furusawa}, {Morokuma}, {Ishida},
  {Aoki}, {Fuse}, {Imanishi}, {Iye}, {Karoji}, {Kobayashi}, {Kodama},
  {Komiyama}, {Maeda}, {Miyazaki}, {Mizumoto}, {Nakata}, {Noumaru},
  {Ogasawara}, {Okamura}, {Saito}, {Sasaki}, {Ueda}, \&
  {Yoshida}}]{Furusawa2008}
{Furusawa} H. {et~al.}, 2008, \apjs, 176, 1

\bibitem[{{Galametz} {et~al}\mbox{.}(2012){Galametz}, {Stern}, {De Breuck},
  {Hatch}, {Mayo}, {Miley}, {Rettura}, {Seymour}, {Stanford}, \&
  {Vernet}}]{Galametz2012}
{Galametz} A. {et~al.}, 2012, \apj, 749, 169

\bibitem[{{Galametz} {et~al}\mbox{.}(2010){Galametz}, {Stern}, {Stanford}, {De
  Breuck}, {Vernet}, {Griffith}, \& {Harrison}}]{Galametz2010a}
{Galametz} A., {Stern} D., {Stanford} S.~A., {De Breuck} C., {Vernet} J.,
  {Griffith} R.~L., {Harrison} F.~A., 2010, \aap, 516, A101

\bibitem[{{Hartley} {et~al}\mbox{.}(2013){Hartley}, {Almaini}, {Mortlock},
  {Conselice}, {Gr{\"u}tzbauch}, {Simpson}, {Bradshaw}, {Chuter}, {Foucaud},
  {Cirasuolo}, {Dunlop}, {McLure}, \& {Pearce}}]{Hartley2013}
{Hartley} W.~G. {et~al.}, 2013, \mnras, 431, 3045

\bibitem[{{Hasselfield} {et~al}\mbox{.}(2013){Hasselfield}, {Hilton},
  {Marriage}, {Addison}, {Barrientos}, {Battaglia}, {Battistelli}, {Bond},
  {Crichton}, {Das}, {Devlin}, {Dicker}, {Dunkley}, {D{\"u}nner}, {Fowler},
  {Gralla}, {Hajian}, {Halpern}, {Hincks}, {Hlozek}, {Hughes}, {Infante},
  {Irwin}, {Kosowsky}, {Marsden}, {Menanteau}, {Moodley}, {Niemack}, {Nolta},
  {Page}, {Partridge}, {Reese}, {Schmitt}, {Sehgal}, {Sherwin}, {Sievers},
  {Sif{\'o}n}, {Spergel}, {Staggs}, {Swetz}, {Switzer}, {Thornton}, {Trac}, \&
  {Wollack}}]{Hasselfield2013}
{Hasselfield} M. {et~al.}, 2013, \jcap, 7, 8

\bibitem[{{Hatch} {et~al}\mbox{.}(2011){Hatch}, {De Breuck}, {Galametz},
  {Miley}, {Overzier}, {R{\"o}ttgering}, {Doherty}, {Kodama}, {Kurk},
  {Seymour}, {Venemans}, {Vernet}, \& {Zirm}}]{Hatch2011a}
{Hatch} N.~A. {et~al.}, 2011, \mnras, 410, 1537

\bibitem[{{Hatch} {et~al}\mbox{.}(2014){Hatch}, {Wylezalek}, {Kurk}, {Stern},
  {De Breuck}, {Jarvis}, {Galametz}, {Gonzalez}, {Hartley}, {Mortlock},
  {Seymour}, \& {Stevens}}]{Hatch2014}
{Hatch} N.~A. {et~al.}, 2014, \mnras, 445, 280

\bibitem[{{Kodama} {et~al}\mbox{.}(2007){Kodama}, {Tanaka}, {Kajisawa}, {Kurk},
  {Venemans}, {De Breuck}, {Vernet}, \& {Lidman}}]{Kodama2007}
{Kodama} T., {Tanaka} I., {Kajisawa} M., {Kurk} J., {Venemans} B., {De Breuck}
  C., {Vernet} J., {Lidman} C., 2007, \mnras, 377, 1717

\bibitem[{{Kriek} {et~al}\mbox{.}(2008){Kriek}, {van Dokkum}, {Franx},
  {Illingworth}, {Marchesini}, {Quadri}, {Rudnick}, {Taylor}, {F{\"o}rster
  Schreiber}, {Gawiser}, {Labb{\'e}}, {Lira}, \& {Wuyts}}]{Kriek2008}
{Kriek} M. {et~al.}, 2008, \apj, 677, 219

\bibitem[{{Lacy} {et~al}\mbox{.}(1999){Lacy}, {Rawlings}, {Hill}, {Bunker},
  {Ridgway}, \& {Stern}}]{Lacy1999}
{Lacy} M., {Rawlings} S., {Hill} G.~J., {Bunker} A.~J., {Ridgway} S.~E.,
  {Stern} D., 1999, \mnras, 308, 1096

\bibitem[{{Lee} {et~al}\mbox{.}(2015){Lee}, {Im}, {Kim}, {Lotz}, {McPartland},
  {Peth}, \& {Koekemoer}}]{Lee2015}
{Lee} S.-K., {Im} M., {Kim} J.-W., {Lotz} J., {McPartland} C., {Peth} M.,
  {Koekemoer} A., 2015, ArXiv e-prints

\bibitem[{{Mortlock} {et~al}\mbox{.}(2013){Mortlock}, {Conselice}, {Hartley},
  {Ownsworth}, {Lani}, {Bluck}, {Almaini}, {Duncan}, {Wel}, {Koekemoer},
  {Dekel}, {Dav{\'e}}, {Ferguson}, {de Mello}, {Newman}, {Faber}, {Grogin},
  {Kocevski}, \& {Lai}}]{Mortlock2013}
{Mortlock} A. {et~al.}, 2013, \mnras, 433, 1185

\bibitem[{{Muldrew}, {Hatch} \& {Cooke}(2015){Muldrew}, {Hatch}, \&
  {Cooke}}]{Muldrew2015}
{Muldrew} S.~I., {Hatch} N.~A., {Cooke} E.~A., 2015, \mnras, 452, 2528

\bibitem[{{Muzzin} {et~al}\mbox{.}(2009){Muzzin}, {Wilson}, {Yee}, {Hoekstra},
  {Gilbank}, {Surace}, {Lacy}, {Blindert}, {Majumdar}, {Demarco}, {Gardner},
  {Gladders}, \& {Lonsdale}}]{Muzzin2009}
{Muzzin} A. {et~al.}, 2009, \apj, 698, 1934

\bibitem[{{Newman} {et~al}\mbox{.}(2014){Newman}, {Ellis}, {Andreon}, {Treu},
  {Raichoor}, \& {Trinchieri}}]{Newman2014}
{Newman} A.~B., {Ellis} R.~S., {Andreon} S., {Treu} T., {Raichoor} A.,
  {Trinchieri} G., 2014, \apj, 788, 51

\bibitem[{{Orsi} {et~al}\mbox{.}(2015){Orsi}, {Fanidakis}, {Lacey}, \&
  {Baugh}}]{Orsi2015}
{Orsi} A.~A., {Fanidakis} N., {Lacey} C.~G., {Baugh} C.~M., 2015, ArXiv
  e-prints

\bibitem[{{Overzier} {et~al}\mbox{.}(2005){Overzier}, {Harris}, {Carilli},
  {Pentericci}, {R{\"o}ttgering}, \& {Miley}}]{Overzier2005}
{Overzier} R.~A., {Harris} D.~E., {Carilli} C.~L., {Pentericci} L.,
  {R{\"o}ttgering} H.~J.~A., {Miley} G.~K., 2005, \aap, 433, 87

\bibitem[{{Papovich}(2008)}]{Papovich2008}
{Papovich} C., 2008, \apj, 676, 206

\bibitem[{{Papovich} {et~al}\mbox{.}(2012){Papovich}, {Bassett}, {Lotz}, {van
  der Wel}, {Tran}, {Finkelstein}, {Bell}, {Conselice}, {Dekel}, {Dunlop},
  {Guo}, {Faber}, {Farrah}, {Ferguson}, {Finkelstein}, {H{\"a}ussler},
  {Kocevski}, {Koekemoer}, {Koo}, {McGrath}, {McLure}, {McIntosh}, {Momcheva},
  {Newman}, {Rudnick}, {Weiner}, {Willmer}, \& {Wuyts}}]{Papovich2012}
{Papovich} C. {et~al.}, 2012, \apj, 750, 93

\bibitem[{{Papovich} {et~al}\mbox{.}(2010){Papovich}, {Momcheva}, {Willmer},
  {Finkelstein}, {Finkelstein}, {Tran}, {Brodwin}, {Dunlop}, {Farrah}, {Khan},
  {Lotz}, {McCarthy}, {McLure}, {Rieke}, {Rudnick}, {Sivanandam}, {Pacaud}, \&
  {Pierre}}]{Papovich2010}
{Papovich} C. {et~al.}, 2010, \apj, 716, 1503

\bibitem[{{Pierre} {et~al}\mbox{.}(2012){Pierre}, {Clerc}, {Maughan}, {Pacaud},
  {Papovich}, \& {Willmer}}]{Pierre2012}
{Pierre} M., {Clerc} N., {Maughan} B., {Pacaud} F., {Papovich} C., {Willmer}
  C.~N.~A., 2012, \aap, 540, A4

\bibitem[{{Planck Collaboration} {et~al}\mbox{.}(2015){Planck Collaboration},
  {Ade}, {Aghanim}, {Arnaud}, {Ashdown}, {Aumont}, {Baccigalupi}, {Banday},
  {Barreiro}, {Barrena}, \& et~al.}]{Planck2015}
{Planck Collaboration} {et~al.}, 2015, ArXiv e-prints

\bibitem[{{Rettura} {et~al}\mbox{.}(2014){Rettura}, {Martinez-Manso}, {Stern},
  {Mei}, {Ashby}, {Brodwin}, {Gettings}, {Gonzalez}, {Stanford}, \&
  {Bartlett}}]{Rettura2014}
{Rettura} A. {et~al.}, 2014, \apj, 797, 109

\bibitem[{{Rudnick} {et~al}\mbox{.}(2012){Rudnick}, {Tran}, {Papovich},
  {Momcheva}, \& {Willmer}}]{Rudnick2012}
{Rudnick} G.~H., {Tran} K.-V., {Papovich} C., {Momcheva} I., {Willmer} C.,
  2012, \apj, 755, 14

\bibitem[{{Santos} {et~al}\mbox{.}(2015){Santos}, {Altieri}, {Valtchanov},
  {Nastasi}, {B{\"o}hringer}, {Cresci}, {Elbaz}, {Fassbender}, {Rosati},
  {Tozzi}, \& {Verdugo}}]{Santos2015}
{Santos} J.~S. {et~al.}, 2015, \mnras, 447, L65

\bibitem[{{Schirmer}(2013)}]{Schirmer2013}
{Schirmer} M., 2013, \apjs, 209, 21

\bibitem[{{Simpson} \& {Eisenhardt}(1999)}]{Simpson1999}
{Simpson} C., {Eisenhardt} P., 1999, \pasp, 111, 691

\bibitem[{{Simpson} \& {Rawlings}(2002)}]{Simpson2002}
{Simpson} C., {Rawlings} S., 2002, \mnras, 334, 511

\bibitem[{{Smail} {et~al}\mbox{.}(2014){Smail}, {Geach}, {Swinbank}, {Tadaki},
  {Arumugam}, {Hartley}, {Almaini}, {Bremer}, {Chapin}, {Chapman}, {Danielson},
  {Edge}, {Scott}, {Simpson}, {Simpson}, {Conselice}, {Dunlop}, {Ivison},
  {Karim}, {Kodama}, {Mortlock}, {Robson}, {Roseboom}, {Thomson}, {van der
  Werf}, \& {Webb}}]{Smail2014}
{Smail} I. {et~al.}, 2014, \apj, 782, 19

\bibitem[{{Snyder} {et~al}\mbox{.}(2012){Snyder}, {Brodwin}, {Mancone},
  {Zeimann}, {Stanford}, {Gonzalez}, {Stern}, {Eisenhardt}, {Brown}, {Dey},
  {Jannuzi}, \& {Perlmutter}}]{Snyder2012}
{Snyder} G.~F. {et~al.}, 2012, \apj, 756, 114

\bibitem[{{Stanford} {et~al}\mbox{.}(2012){Stanford}, {Brodwin}, {Gonzalez},
  {Zeimann}, {Stern}, {Dey}, {Eisenhardt}, {Snyder}, \&
  {Mancone}}]{Stanford2012}
{Stanford} S.~A. {et~al.}, 2012, \apj, 753, 164

\bibitem[{{Stanford} {et~al}\mbox{.}(2005){Stanford}, {Eisenhardt}, {Brodwin},
  {Gonzalez}, {Stern}, {Jannuzi}, {Dey}, {Brown}, {McKenzie}, \&
  {Elston}}]{Stanford2005}
{Stanford} S.~A. {et~al.}, 2005, \apjl, 634, L129

\bibitem[{{Stanford} {et~al}\mbox{.}(2014){Stanford}, {Gonzalez}, {Brodwin},
  {Gettings}, {Eisenhardt}, {Stern}, \& {Wylezalek}}]{Stanford2014}
{Stanford} S.~A., {Gonzalez} A.~H., {Brodwin} M., {Gettings} D.~P.,
  {Eisenhardt} P.~R.~M., {Stern} D., {Wylezalek} D., 2014, \apjs, 213, 25

\bibitem[{{Tanaka}, {Finoguenov} \& {Ueda}(2010){Tanaka}, {Finoguenov}, \&
  {Ueda}}]{Tanaka2010}
{Tanaka} M., {Finoguenov} A., {Ueda} Y., 2010, \apjl, 716, L152

\bibitem[{{Tozzi} {et~al}\mbox{.}(2015){Tozzi}, {Santos}, {Jee}, {Fassbender},
  {Rosati}, {Nastasi}, {Forman}, {Sartoris}, {Borgani}, {Boehringer},
  {Altieri}, {Pratt}, {Nonino}, \& {Jones}}]{Tozzi2015}
{Tozzi} P. {et~al.}, 2015, \apj, 799, 93

\bibitem[{{Tran} {et~al}\mbox{.}(2015){Tran}, {Nanayakkara}, {Yuan},
  {Kacprzak}, {Glazebrook}, {Kewley}, {Momcheva}, {Papovich}, {Quadri},
  {Rudnick}, {Saintonge}, {Spitler}, {Straatman}, \& {Tomczak}}]{Tran2015}
{Tran} K.-V.~H. {et~al.}, 2015, ArXiv e-prints

\bibitem[{{van Breukelen} {et~al}\mbox{.}(2009){van Breukelen}, {Simpson},
  {Rawlings}, {Akiyama}, {Bonfield}, {Clewley}, {Jarvis}, {Mauch}, {Readhead},
  {Stobbart}, {Swinbank}, \& {Watson}}]{vanBreukelen2009}
{van Breukelen} C. {et~al.}, 2009, \mnras, 395, 11

\bibitem[{{van der Burg} {et~al}\mbox{.}(2013){van der Burg}, {Muzzin},
  {Hoekstra}, {Lidman}, {Rettura}, {Wilson}, {Yee}, {Hildebrandt},
  {Marchesini}, {Stefanon}, {Demarco}, \& {Kuijken}}]{VanderBurg2013}
{van der Burg} R.~F.~J. {et~al.}, 2013, \aap, 557, A15

\bibitem[{{Venemans} {et~al}\mbox{.}(2007){Venemans}, {R{\"o}ttgering},
  {Miley}, {van Breugel}, {de Breuck}, {Kurk}, {Pentericci}, {Stanford},
  {Overzier}, {Croft}, \& {Ford}}]{Venemans2007}
{Venemans} B.~P. {et~al.}, 2007, \aap, 461, 823

\bibitem[{{Williams} {et~al}\mbox{.}(2009){Williams}, {Quadri}, {Franx}, {van
  Dokkum}, \& {Labb{\'e}}}]{Williams2009a}
{Williams} R.~J., {Quadri} R.~F., {Franx} M., {van Dokkum} P., {Labb{\'e}} I.,
  2009, \apj, 691, 1879

\bibitem[{{Willis} {et~al}\mbox{.}(2013){Willis}, {Clerc}, {Bremer}, {Pierre},
  {Adami}, {Ilbert}, {Maughan}, {Maurogordato}, {Pacaud}, {Valtchanov},
  {Chiappetti}, {Thanjavur}, {Gwyn}, {Stanway}, \& {Winkworth}}]{Willis2013}
{Willis} J.~P. {et~al.}, 2013, \mnras, 430, 134

\bibitem[{{Wylezalek} {et~al}\mbox{.}(2013){Wylezalek}, {Galametz}, {Stern},
  {Vernet}, {De Breuck}, {Seymour}, {Brodwin}, {Eisenhardt}, {Gonzalez},
  {Hatch}, {Jarvis}, {Rettura}, {Stanford}, \& {Stevens}}]{Wylezalek2013}
{Wylezalek} D. {et~al.}, 2013, \apj, 769, 79

\bibitem[{{Wylezalek} {et~al}\mbox{.}(2014){Wylezalek}, {Vernet}, {De Breuck},
  {Stern}, {Brodwin}, {Galametz}, {Gonzalez}, {Jarvis}, {Hatch}, {Seymour}, \&
  {Stanford}}]{Wylezalek2014}
{Wylezalek} D. {et~al.}, 2014, \apj, 786, 17

\end{thebibliography}


\end{document}